\documentclass[sigconf]{acmart}
\usepackage[T1]{fontenc}
\usepackage[utf8]{inputenc}
\usepackage[english]{babel}
\usepackage{nicefrac}
\usepackage{csquotes}
\usepackage{mathtools,amsthm}
\usepackage{booktabs}
\usepackage{enumitem}
\usepackage[binary-units=true]{siunitx}
\usepackage{csvsimple}
\usepackage{tikz}
\usepackage{breakurl}

\usetikzlibrary{backgrounds,fit,decorations.pathreplacing,calc,matrix,arrows.meta}
\newtheorem{example}{Example}

\newcommand{\ket}[1]{\ensuremath{\left|#1\right\rangle}}
\hypersetup{pdfinfo={Author={Stefan Hillmich, Alwin Zulehner, and Robert Wille}}}
\renewcommand\footnotetextcopyrightpermission[1]{}
\copyrightyear{2021}
\acmYear{2021}
\setcopyright{acmcopyright}
\acmConference[ASPDAC '21]{26th Asia and South Pacific Design Automation Conference}{January 18--21, 2021}{Tokyo, Japan}
\acmBooktitle{26th Asia and South Pacific Design Automation Conference (ASPDAC '21), January 18--21, 2021, Tokyo, Japan}
\acmPrice{15.00}
\acmDOI{10.1145/3394885.3431604}
\acmISBN{978-1-4503-7999-1/21/01}
\renewcommand{\baselinestretch}{0.94}
\begin{document}
\title{Exploiting Quantum Teleportation in Quantum Circuit Mapping}
\date{}
\author[Stefan Hillmich, Alwin Zulehner, and Robert Wille]{\vspace{-1em}Stefan Hillmich$^*$\hspace{3.0em}Alwin Zulehner$^*$\hspace{3.0em}Robert Wille$^{*\dagger}$}
\affiliation{%
   \institution{\vspace{0.5em}$^*$Johannes Kepler University Linz, Austria}%
}
\affiliation{%
  \institution{$^\dagger$Software Competence Center Hagenberg GmbH (SCCH), Austria}
}
\email{{stefan.hillmich, robert.wille}@jku.at}
\email{https://iic.jku.at/eda/research/quantum/}

\begin{abstract}
	Quantum computers are constantly growing in their number of qubits, but continue to suffer from restrictions such as the limited pairs of qubits that may interact with each other.
	Thus far, this problem is addressed by mapping and moving qubits to suitable positions for the interaction (known as \emph{quantum circuit mapping}).
	However, this movement requires additional gates to be incorporated into the circuit, whose number should be kept as small as possible since each gate increases the likelihood of errors and decoherence.
	State-of-the-art mapping methods utilize swapping and bridging to move the qubits along the static paths of the coupling map---solving this problem without exploiting all means the quantum domain has to offer. 
	In this paper, we propose to additionally exploit quantum teleportation as a possible complementary method. 
	Quantum teleportation conceptually allows to move the state of a qubit over arbitrary long distances with constant overhead---providing the potential of determining cheaper mappings.	
	The potential is demonstrated by a case study on the IBM Q Tokyo architecture which already shows promising improvements. 
	With the emergence of larger quantum computing architectures, quantum teleportation will become more effective in generating cheaper mappings.
\end{abstract}
\maketitle

\section{Introduction}
\label{sec:introduction}

Quantum computing~\cite{NC:2000} enables significant improvements over classical computing for certain problems and is providing an exponential speedup in the best case.
Well known examples for such problems are integer factorization using Shor's algorithm~\cite{Sho:94}, quantum chemistry~\cite{cao2019quantum}, and boson sampling~\cite{DBLP:conf/soda/CliffordC18}.
The commonly used description for quantum algorithms are quantum circuits which represent a series of operations to be performed on the quantum state. 
However, physical realizations of current quantum computers are considered \emph{Noisy Intermediate Scale Quantum} (NISQ~\cite{preskill2018quantum}) devices and they impose restrictions that have to be explicitly addressed in the quantum circuit descriptions before they can be executed on a physical quantum computer.
\newpage

A particularly important restriction is given by the \emph{coupling constraint} that only allows interactions between specific pairs of qubits. 
In all but the trivial cases, it is not possible to map the logical qubits of a quantum circuit to the physical qubits of a quantum computer in a way that the coupling constraints are satisfied for the whole circuit. 
This is a problem solved through \emph{quantum circuit mapping}\footnote{The quantum circuit mapping is commonly performed as part of a compilation or synthesis procedure, that handles all restrictions of the targeted quantum computer.}, which moves logical qubits of the circuit diagram to physical qubit positions of the hardware that allow for the desired interactions.
It has been shown that finding an optimal quantum circuit mapping is an \(\mathsf{NP}\)-complete problem~\cite{siraichi2018qubit,botea2018compiling}.

State-of-the-art approaches (such as introduced in~\cite{DBLP:journals/tcad/WilleLD14,DBLP:conf/dac/WilleBZ19,DBLP:conf/aspdac/ItokoRIMC19,SWD:2010b,DBLP:conf/aspdac/WilleKWRCD16,DBLP:journals/tcad/ZulehnerPW19,zulehner2019compiling,8702439,DBLP:journals/corr/RahmanD15,dueck2018optimization,de2019cnot,DBLP:conf/rc/HattoriY18,DBLP:conf/rc/MatsuoY19,2019arXiv190804226L,DBLP:journals/integration/ItokoRIM20,li2020qubits}) employ \emph{swapping} as well as \emph{bridging} schemes to satisfy the coupling constraint for a given operation if required.
While they are working reasonably well for smaller architectures, they lead to substantial additional costs when large distances in growing architectures have to be covered.
Moreover, swapping and bridging are rather classical approaches compared to the untapped potential of the available \mbox{quantum-mechanical} phenomena.

In this work, we are aiming to broaden the consideration of solutions for quantum circuit mapping by additionally exploiting \emph{quantum teleportation}~\cite{NC:2000}---a quantum mechanical method which allows to teleport the state of a qubit over an arbitrary distance through a quantum transportation channel.
Quantum teleportation has been demonstrated for quantum networks~\cite{sun2016quantum,valivarthi2016quantum,valivarthi2020teleportation} but this concept also allows to move logical qubits around inside a quantum computer~\cite{Gottesman_1999,DBLP:conf/tqc/Rosenbaum13,DBLP:journals/qic/PhamS13} and, by this, potentially helps to satisfy the coupling constraints. 
Moreover, moving logic qubits by teleportation via established transportation channels can be accomplished with constant costs, i.e.,~it does not depend at all on the distance to be covered. 
Motivated by this, we propose to additionally exploit quantum teleportation in quantum circuit mapping and also show that this natively fits into currently developed NISQ devices.

A case study conducted on the IBM Q Tokyo architecture demonstrates that this approach results in significant improvements for current state-of-the-art approaches. 
Moreover, the proposed idea will have an amplified impact on larger quantum computing architectures (which already have been announced, e.g., by IBM and Google), since larger distances between qubits have to be considered on these architectures. 
By this, the proposed idea is going to become more effective with growing architectures.

The remainder of the paper is structured as follows:
In \autoref{sec:background}, we motivate the problem including a brief recapitulation of the relevant basics of quantum computations as well as the architectural constraints. This section also reviews the state of the art regarding the mapping of quantum circuits.
Afterwards, \autoref{sec:general-idea} reviews quantum teleportation and introduces the idea of incorporating this phenomenon into the mapping process.
To this end, \autoref{sec:case-study} provides a case study of the impact of quantum teleportation in the mapping process for the 20-qubit IBM Q Tokyo architecture.
Finally, the paper is concluded in \autoref{sec:conclusions}.

\section{Background \& State of the Art}
\label{sec:background}

In this section, we review the problem of quantum circuit mapping considered in this paper and discuss the \mbox{state-of-the-art} solutions which have been introduced thus far to tackle this problem.

\subsection{Considered Problem}
\label{subsec:quantum-circuits}

Quantum circuits are means of describing operations on \emph{qubits}~\cite{NC:2000}.
These operations can act on single or multiple qubits---although without loss of generality we restrict multi-qubit operations to the controlled-\(\mathit{NOT}\) (\(\mathit{CNOT}\)) operation in this work.
We further distinguish between control qubits and target qubits, where the operation is performed on the target qubit if and only if the control qubit assumes the state \ket{1}.\footnote{The precise functionality of the respective operations is irrelevant in this work and, hence, is omitted. However, for a more detailed treatment, we refer the interested reader to~\cite{NC:2000}.}

The graphical representation of quantum circuits uses horizontal lines to denote the qubits, which pass through gates (representing operations) that manipulate the qubits.
This may seem similar to circuits in the classical realm, however, the circuit just describes the order (from left to right) in which the gates/operations are applied to the qubits.

\begin{figure}
	\centering
	\begin{tikzpicture}[gate/.style={draw,fill=white,minimum size=1.5em},
		control/.style={draw,fill,shape=circle,minimum size=5pt,inner sep=0pt},
		cross/.style={path picture={\draw[thick,black](path picture bounding box.north) -- (path picture bounding box.south) (path picture bounding box.west) -- (path picture bounding box.east);}},
		target/.style={draw,circle,cross,minimum width=0.3cm},
		x=0.75cm, y=0.75cm]
		\node[anchor=east] (q0) at (0,2) {\(q_0=\ket{0}\)};
		\node[anchor=east] (q1) at (0,1) {\(q_1=\ket{0}\)};
		\node[anchor=east] (q2) at (0,0) {\(q_2=\ket{0}\)};
		
		\node[gate] at (1, |-q1) {\(H\)};
		\node[gate] at (2, |-q0) {\(X\)};
		
		\draw (3,|-q1) node[control] {} -- (3,|-q2) node [target] {};
		\draw (4,|-q0) node[control] {} -- (4,|-q2) node [target] {};
		\draw (5,|-q1) node[control] {} -- (5,|-q0) node [target] {};
		\node[gate] at (6, |-q0) {\(T\)};
		
		\begin{scope}[on background layer]
			\draw (0,0) -- (6.75,0);
			\draw (0,1) -- (6.75,1);
			\draw (0,2) -- (6.75,2);
		\end{scope}
	\end{tikzpicture}
	\vspace{-1em}
	\caption{Quantum circuit diagram}
	\label{fig:quantum-circuit}
\end{figure}
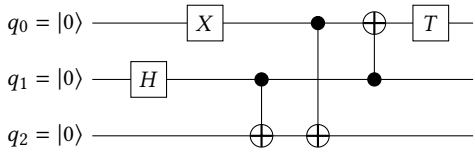

\begin{example}
	\newcommand{\control}{\tikz[control/.style={draw,fill,shape=circle,minimum size=5pt,inner sep=0pt}]{\node[control] {};}}
	\newcommand{\target}{\tikz[cross/.style={path picture={\draw[thick,black](path picture bounding box.north) -- (path picture bounding box.south) (path picture bounding box.west) -- (path picture bounding box.east);}}]{\node[draw,circle,cross,scale=0.9] {};}}
	\autoref{fig:quantum-circuit} depicts the diagram of a quantum circuit. 
	It is composed of three qubits and six gates.
	The gates marked with \(H\), \(X\), and \(T\) are single-qubit gates.
	For the multi-qubit \(\mathit{CNOT}\) gates, the control qubit is represented by\enspace\control\ whereas the target qubit is represented by\enspace\target.
	In the presented diagram, \(H\) and \(X\) are applied on the first two qubits, respectively, followed by three \(\mathit{CNOT}\) gates and, finally, a single \(T\) gate.
\end{example}

Quantum circuit diagrams are commonly  agnostic to physical architectures, i.e.,~they focus on the functionality without addressing physical restrictions of physical quantum computers. In fact, physical quantum architectures may only be able to apply certain quantum operations or limit the possible interactions between qubits.
The restriction of interaction is referred to as \emph{coupling constraint} and is a main focus of this paper.

Due to the coupling constraint, \(\mathit{CNOT}\) operations cannot be applied between arbitrary pairs of physical qubits of the quantum computer.
The possible pairs are defined in the \emph{coupling~map}---commonly depicted as a graph with nodes for the qubits and edges denoting possible \(\mathit{CNOT}\) positions.
Here, physical qubits  are usually denoted by~\(Q_i\) in comparison to the logical qubits which are denoted by~\(q_i\).
As single-qubit gates do not interact directly with other qubits, they are unaffected by the coupling constraint.

\begin{figure}[!tp]
	\centering
	\scalebox{0.8}{\begin{tikzpicture}[thick, qubit/.style={draw, circle, minimum size=2.75em, fill=white}]
		\matrix (tokyo) [matrix of math nodes, ampersand replacement=\&, every node/.append style={qubit}, row sep=0.5cm, column sep=1cm] {
			Q_0 \& Q_1 \& Q_2 \& Q_3 \& Q_4 \\
			Q_5 \& Q_6 \& Q_7 \& Q_8 \& Q_9 \\
			Q_{10} \& Q_{11} \& Q_{12} \& Q_{13} \& Q_{14} \\
			Q_{15} \& Q_{16} \& Q_{17} \& Q_{18} \& Q_{19} \\
		};
		\begin{scope}[on background layer, every path/.style={thick}]
		\draw (tokyo-1-1) -- (tokyo-1-5);
		\draw (tokyo-2-1) -- (tokyo-2-5);
		\draw (tokyo-3-1) -- (tokyo-3-5);
		\draw (tokyo-4-1) -- (tokyo-4-5);
		
		\draw (tokyo-1-1) -- (tokyo-4-1);
		\draw (tokyo-1-2) -- (tokyo-4-2);
		\draw (tokyo-1-3) -- (tokyo-4-3);
		\draw (tokyo-1-4) -- (tokyo-4-4);
		\draw (tokyo-1-5) -- (tokyo-4-5);
		
		\draw (tokyo-1-2) -- (tokyo-4-5);
		\draw (tokyo-1-3) -- (tokyo-3-1);
		\draw (tokyo-1-4) -- (tokyo-2-5);
		\draw (tokyo-1-5) -- (tokyo-4-2);
		\draw (tokyo-2-1) -- (tokyo-4-3);
		\draw (tokyo-3-5) -- (tokyo-4-4);
		\end{scope}
		\end{tikzpicture}}
	\vspace{-1em}
	\caption{Coupling map for the IBM Q Tokyo architecture}
	\label{fig:ibm-tokyo}
\end{figure}
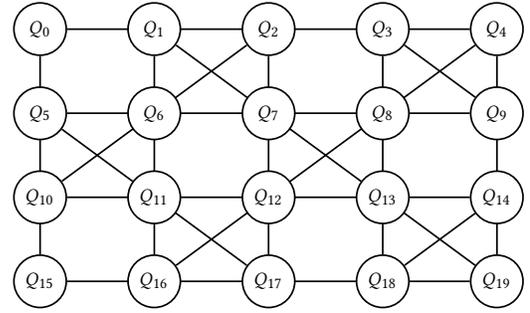

\begin{example}
	The quantum circuit depicted in \autoref{fig:quantum-circuit} is architecture-agnostic with logical qubits labeled \(q_0\), \(q_1\), and \(q_2\).
	Further, \autoref{fig:ibm-tokyo} shows a coupling map of the \emph{IBM Q Tokyo} architecture, IBM's 20 qubit NISQ device, where \(\mathit{CNOT}\) gates can only be applied between physical qubits \(Q_i\) and \(Q_j\) that are connected by an edge in the coupling map (e.g.,~\(Q_0\) and \(Q_5\)).
\end{example}

The restricted interactions lead to the problem of how to satisfy the coupling constraint for arbitrary circuits with an as small as possible number of additional gates, i.e.,~how to determine an \emph{efficient quantum circuit mapping}. 

\begin{example}\label{ex:cc-violated}
	Consider again the circuit in \autoref{fig:quantum-circuit} with the coupling map of the IBM Q Tokyo architecture as shown in \autoref{fig:ibm-tokyo}.	Furthermore, assume the mapping puts the logical qubits on the physical ones with the same index, i.e.,~\(Q_i \leftarrow q_i \).
	In this case, the second \(\mathit{CNOT}\) gate in \autoref{fig:quantum-circuit} cannot be applied since the coupling constraints are not satisfied since there is no connection between \(Q_0\) and \(Q_2\).
\end{example}

\subsection{State of the Art \& Limitations}
\label{subsec:classical}

The architectural constraints in current NISQ quantum computers do not allow applying \mbox{two-qubit} gates, e.g.,~\(\mathit{CNOT}\) gates, between arbitrary qubits, but only for specific pairs of qubits as specified by the corresponding coupling map of the architecture.
Since determining a mapping of logical qubits to physical qubits satisfying the constraint for \emph{all} gates of the circuit is only possible in trivial cases, additional quantum operations need to be inserted into the circuit to satisfy the coupling constraint.
This increases the gate count and, by this, the cost and unreliability of the circuit (quantum computers employ error-rates for gate operations in the range of \(10^{-3}\)~\cite{qxbackends2019}).
The goal is to add as few additional quantum operations as possible---an \mbox{\(\mathsf{NP}\)-complete} problem for the exact solution~\cite{botea2018compiling,siraichi2018qubit}.

In the past, several methods have been developed that tackle this problem~\cite{DBLP:journals/tcad/WilleLD14,DBLP:conf/dac/WilleBZ19,DBLP:conf/aspdac/ItokoRIMC19,SWD:2010b,DBLP:conf/aspdac/WilleKWRCD16,DBLP:journals/tcad/ZulehnerPW19,zulehner2019compiling,8702439,DBLP:journals/corr/RahmanD15,dueck2018optimization,de2019cnot,DBLP:conf/rc/HattoriY18,DBLP:conf/rc/MatsuoY19,2019arXiv190804226L,DBLP:journals/integration/ItokoRIM20,li2020qubits}. 
All these algorithms follow one of two strategies for satisfying the coupling constraint: using \emph{swapping} or \emph{bridging}.

\begin{figure}
	\centering
	\resizebox{\linewidth}{!}{\begin{tikzpicture}[gate/.style={draw,fill=white,minimum size=1.5em},
		control/.style={draw,fill,shape=circle,minimum size=5pt,inner sep=0pt},
		cross/.style={path picture={\draw[thick,black](path picture bounding box.north) -- (path picture bounding box.south) (path picture bounding box.west) -- (path picture bounding box.east);}},
		target/.style={draw,circle,cross,minimum width=0.3cm},
		swap/.style={path picture={\draw[thick,black](path picture bounding box.north east) -- (path picture bounding box.south west) (path picture bounding box.north west) -- (path picture bounding box.south east);}},
		x=0.75cm, y=0.75cm]
		
		\begin{scope}[xshift=-0.5cm]
			\node[anchor=east] (q0) at (0,2) {\(q_0\)};
			\node[anchor=east] (q1) at (0,1) {\(q_1\)};
			\node[anchor=east] (q2) at (0,0) {\(q_2\)};
			
			\draw (0.5,|-q0) node[control] {} -- (.5,|-q2) node [target] {};
		
			\begin{scope}[on background layer]
				\draw (0,0) -- (1,0);
				\draw (0,1) -- (1,1);
				\draw (0,2) -- (1,2);
			\end{scope}
			
			\node at (1.5,1) {\(\to\)};
		\end{scope}

		\node[anchor=east] (q0b) at (3,2) {\(Q_0 \leftarrow q_0\)};
		\node[anchor=east] (q1b) at (3,1) {\(Q_1 \leftarrow q_1\)};
		\node[anchor=east] (q2b) at (3,0) {\(Q_2 \leftarrow q_2\)};
		
		\draw (3.5,|-q0b) node[swap] {} -- (3.5,|-q1b) node [swap] {};
		\draw (4.5,|-q1b) node[control] {} -- (4.5,|-q2b) node [target] {};
		
		\begin{scope}[on background layer]
		\draw (3,0) -- ++(2,0) node[right] {\(q_2\)};
		\draw (3,1) -- ++(2,0) node[right] {\(q_0\)};
		\draw (3,2) -- ++(2,0) node[right] {\(q_1\)};
		\end{scope}

		\node at (6,1) {\(\to\)};
		
		\node[anchor=east] (q0c) at (8,2) {\(Q_0 \leftarrow q_0\)};
		\node[anchor=east] (q1c) at (8,1) {\(Q_1 \leftarrow q_1\)};
		\node[anchor=east] (q2c) at (8,0) {\(Q_2 \leftarrow q_2\)};
		
		\draw (8.5,|-q0c) node[control] {} -- (8.5,|-q1c) node [target] {};
		\draw (9.5,|-q1c) node[control] {} -- (9.5,|-q0c) node [target] {};
		\draw (10.5,|-q0c) node[control] {} -- (10.5,|-q1c) node [target] {};
		\draw (11.5,|-q1c) node[control] {} -- (11.5,|-q2c) node [target] {};
		
		\draw [decorate,decoration={brace,amplitude=6pt,raise=8pt}] (8.25,|-q0c) -- (10.75,|-q0c) node [black,midway,above, yshift=1.25em,align=center,font=\scriptsize] {\(\mathit{SWAP}\) decomposition};
		
		\begin{scope}[on background layer]
		\draw (8,0) -- ++(4,0) node[right] {\(q_2\)};
		\draw (8,1) -- ++(4,0) node[right] {\(q_0\)};
		\draw (8,2) -- ++(4,0) node[right] {\(q_1\)};
		\end{scope}
		\end{tikzpicture}}
	\caption{Mapping and decomposition using \(\mathit{SWAP}\) gates}
	\label{fig:swap-decomposition}
\end{figure}
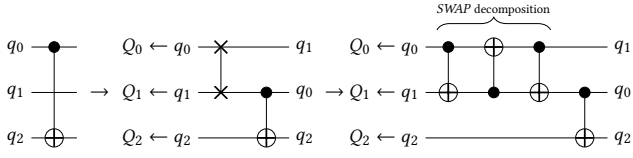

\newcommand{\swap}{\tikz[swap/.style={path picture={\draw[thick,black](path picture bounding box.north east) -- (path picture bounding box.south west) (path picture bounding box.north west) -- (path picture bounding box.south east);}}]{\node[swap, scale=0.9] {};}}
The general idea of \emph{swapping} is to insert \(\mathit{SWAP}\) operations that exchange the state of two physical qubits.
In circuit diagrams the \(\mathit{SWAP}\) is represented by two \swap\ which are connected vertically.
Algorithms utilizing swapping start with an initial mapping of the \(n\) logical qubits \(q_0,q_1,\ldots,q_{n-1}\) to the $m \geq n$ physical qubits \(Q_0,Q_1,\ldots,Q_{m-1}\). 
In case that there is no direct connection in the coupling map for a given \(\mathit{CNOT}\) gate, the target and the control qubit are moved towards each other by inserting \(\mathit{SWAP}\) operations.
By this, the mapping of the logical qubits of the quantum circuit to the physical ones of the hardware changes dynamically, i.e.,~the logical qubits are moved around on the physical ones. 
As swapping is not a native operation in commonly considered architectures, it has to be decomposed.
The minimal decomposition is shown in \autoref{fig:swap-decomposition}.

\begin{example}
	Consider again \autoref{fig:quantum-circuit} and assume the coupling map given in \autoref{fig:ibm-tokyo} with logical qubits \(q_i\) initially mapped to physical qubits \(Q_i\).
	The problem described in \autoref{subsec:quantum-circuits} and illustrated in \autoref{ex:cc-violated} can be addressed by adding a single \(\mathit{SWAP}\) gate placed on \(q_0\) and \(q_1\) after the first \(\mathit{CNOT}\).
	The remaining operations (one \(\mathit{CNOT}\) and one \(T\)) can still be applied with the changed permutation of logical qubits, i.e., no further \(\mathit{SWAP}\) operations are required.
	Naturally, the changed order has to be considered if the circuit is extended or measured, however, reversing the effects of the \(\mathit{SWAP}\) operations is generally not necessary.
\end{example}

In the past, several algorithms have been proposed that follow this general flow, either solving the mapping problem in an exact fashion~\cite{DBLP:journals/tcad/WilleLD14,DBLP:journals/corr/RahmanD15,DBLP:conf/dac/WilleBZ19,DBLP:conf/aspdac/ItokoRIMC19} (which is only feasible for small instances due to the enormous complexity), or by using heuristics~\cite{DBLP:conf/aspdac/ItokoRIMC19,SWD:2010b,DBLP:conf/aspdac/WilleKWRCD16,DBLP:journals/tcad/ZulehnerPW19,zulehner2019compiling,8702439,DBLP:conf/rc/HattoriY18,DBLP:conf/rc/MatsuoY19,2019arXiv190804226L,li2020qubits}.

In contrast to the swapping approach, utilizing \emph{bridging} does not dynamically change the mapping of the logical qubits to the physical ones.
In fact, the initial mapping remains throughout the whole circuit.
\(\mathit{CNOT}\) gates that violate the coupling constraint are realized by a decomposition into several \(\mathit{CNOT}\) gates that bridge the \enquote{gap} in the coupling map, i.e.,~\emph{bridge gates}.

\begin{figure}
	\centering
	\scalebox{0.8}{\begin{tikzpicture}[gate/.style={draw,fill=white,minimum size=1.5em},
		control/.style={draw,fill,shape=circle,minimum size=5pt,inner sep=0pt},
		cross/.style={path picture={\draw[thick,black](path picture bounding box.north) -- (path picture bounding box.south) (path picture bounding box.west) -- (path picture bounding box.east);}},
		target/.style={draw,circle,cross,minimum width=0.3cm},
		x=0.75cm, y=0.75cm]
		\node[anchor=east] (q0) at (0,2) {\(q_0\)};
		\node[anchor=east] (q1) at (0,1) {\(q_1\)};
		\node[anchor=east] (q2) at (0,0) {\(q_2\)};
		
		\draw (0.5,|-q0) node[control] {} -- (.5,|-q2) node [target] {};
		
		\begin{scope}[on background layer]
		\draw (0,0) -- (1,0);
		\draw (0,1) -- (1,1);
		\draw (0,2) -- (1,2);
		\end{scope}
		
		\node at (1.5,1) {\(\to\)};

		\node[anchor=east] (q0b) at (4,2) {\(Q_0 \leftarrow q_0\)};
		\node[anchor=east] (q1b) at (4,1) {\(Q_1 \leftarrow q_1\)};
		\node[anchor=east] (q2b) at (4,0) {\(Q_2 \leftarrow q_2\)};
		
		\draw (4.5,|-q0b) node[control] {} -- (4.5,|-q1b) node [target] {};
		\draw (5.5,|-q1b) node[control] {} -- (5.5,|-q2b) node [target] {};
		\draw (6.5,|-q0b) node[control] {} -- (6.5,|-q1b) node [target] {};
		\draw (7.5,|-q1b) node[control] {} -- (7.5,|-q2b) node [target] {};
		
		\begin{scope}[on background layer]
		\draw (4,0) -- ++(4,0) node[right] {\(q_2\)};
		\draw (4,1) -- ++(4,0) node[right] {\(q_1\)};
		\draw (4,2) -- ++(4,0) node[right] {\(q_0\)};
		\end{scope}
		\end{tikzpicture}}
	\caption{Mapping and decomposition using bridge gates}
	\label{fig:bridge-gates}
\end{figure}
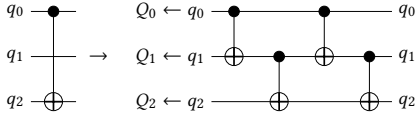

\begin{example}
	Consider again the problem discussed in \autoref{subsec:quantum-circuits} and illustrated in \autoref{ex:cc-violated}.
	Again, the second \(\mathit{CNOT}\) gate violates the coupling constraint.
	Satisfying the constraint using bridge gates is achieved by replacing the offending \(\mathit{CNOT}\) gate with the pattern shown in \autoref{fig:bridge-gates}.
	By this, the permutation of logical qubits remains unchanged but the coupling constraint is satisfied.
\end{example}

The bridging strategy has the advantage that, given the initial mapping, determining the mapped circuit is straightforward. 
Moreover, the cheapest initial (and static) mapping can be determined by counting the interactions between certain qubits. 
In contrast, using swapping hardly allows to determine the cheapest initial mapping since the mapping changes dynamically throughout the circuit and is influenced by many local choices.
The number of \(\mathit{CNOT}\) operations required to realize bridging grows exponentially by \(3\cdot 2^d-2\) gates with respect to the distance \(d\) of the target and the control qubit. 
Several approaches utilizing bridging have been developed~\cite{DBLP:journals/corr/RahmanD15,dueck2018optimization,de2019cnot,DBLP:conf/aspdac/ItokoRIMC19,DBLP:journals/integration/ItokoRIM20}, but typically approaches using swapping result in cheaper mapped circuits since, the overhead with swapping grows only linear with respect to this distance.
Hence, swapping approaches are considered more practical for larger quantum architectures.

\section{Exploiting Quantum Teleportation in Mapping}
\label{sec:general-idea}

As outlined above, state-of-the-art mapping algorithms mainly use swapping for quantum circuit mapping.\footnote{In~\cite{DBLP:conf/tqc/Rosenbaum13,DBLP:journals/qic/PhamS13}, quantum teleportation is used to in a similar way to swapping.}
However, this employs a rather classical way of solving the mapping problem and does not exploit the quantum effects that are available on the target devices anyway.
In this work, we address this by introducing the idea of exploiting \emph{quantum teleportation}~\cite{NC:2000} as a complementary method in mapping. 
Quantum teleportation is well suited to be used in mapping since it allows transferring the state of a qubit over arbitrary distances through a \emph{quantum transportation channel} with a constant overhead by means of quantum operations and a constant classical overhead by means of regular bits to be transferred.

In this section, we show how to extend the swapping approach with quantum teleportation. 
To this end, we first recapitulate the general idea of quantum teleportation. 
This includes a discussion on the feasibility of quantum teleportation on NISQ devices, i.e.,~how all requirements of quantum teleportation are satisfied in the considered setting.

\subsection{Quantum Teleportation}

Quantum teleportation allows to transfer the state of a qubit from one location to another over arbitrary long distances.
Apart from commonly supported quantum operations it only requires transferring two classical bits of information to fully recover the teleported state. 
To this end, two ancillary qubits are required that serve as quantum transportation channel.

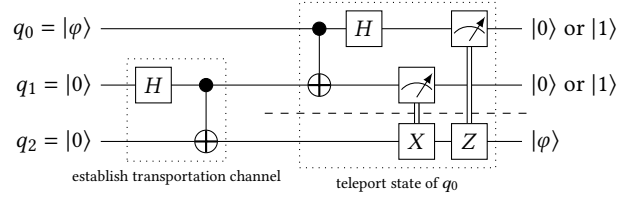
\begin{figure}[!tbp]
	\centering
	\begin{tikzpicture}[gate/.style={draw,fill=white,minimum size=1.5em},
	control/.style={draw,fill,shape=circle,minimum size=5pt,inner sep=0pt},
	cross/.style={fill=white,path picture={\draw[thick,black](path picture bounding box.north) -- (path picture bounding box.south) (path picture bounding box.west) -- (path picture bounding box.east);}},
	target/.style={draw,circle,cross,minimum width=0.3cm},
	classic/.style={double, double distance=1.25pt},
	meter/.append style={
		draw,
		rectangle, 
		minimum size=1.5em, 
		fill=white,
		path picture={
			\draw[black] ([shift={(1pt,5pt)}]path picture bounding box.south west) to[bend left=50] ([shift={(-1pt,5pt)}]path picture bounding box.south east);
			\draw[black,-latex] ([shift={(0,.1)}]path picture bounding box.south) -- ([shift={(.3,-.1)}]path picture bounding box.north);}},
	x=0.7cm, y=0.75cm]
	\node[anchor=east] (q0) at (0,2) {\(q_0 = \ket{\varphi}\)};
	\node[anchor=east] (q1) at (0,1) {\(q_1 = \ket{0}\)};
	\node[anchor=east] (q2) at (0,0) {\(q_2 = \ket{0}\)};
	
	\node[gate] (h1) at (1, |-q1) {\(H\)};
	\draw (2,|-q1) node[control] (c1) {} -- (2,|-q2) node[target] (t1) {};
	\draw[xshift=3pt] (4,|-q0) node[control] (c2) {} -- (4,|-q1) node[target] (t2) {};
	\node[gate] (h2) at (5, |-q0) {\(H\)};
	\draw[classic] (6, |-q1) node[meter](m1) {} -- (6, |-q2) node[gate] (x1) {\(X\)};
	\draw[classic] (7, |-q0) node[meter](m2) {} -- (7, |-q2) node[gate] (z2) {\(Z\)};
	
	\begin{scope}[on background layer]
	\draw (0,|-q2) -- (8,|-q2) node[right] {\(\ket{\varphi}\)};
	\draw (0,|-q1) -- (8,|-q1) node[right] {\(\ket{0}\) or \(\ket{1}\)};
	\draw (0,|-q0) -- (8,|-q0) node[right] {\(\ket{0}\) or \(\ket{1}\)};
	
	\draw[classic] (m1) -- (x1.center);
	\draw[dashed] ($(q1)!0.5!(q2)+(4,0)$) -- ++(5.0,0) node[right,font=\scriptsize] {};
	\end{scope}
	
	\node[draw,dotted,fit=(h1)(c1)(t1),label={[font=\scriptsize]below:establish transportation channel}] {};
	\node[draw,dotted,fit=(c2)(h2)(m2)(t2)(x1),label={[font=\scriptsize]below:teleport state of \(q_0\)}] {};
	\end{tikzpicture}
	\vspace{-0.5em}
	\caption{Circuit diagram for quantum teleportation}
	\label{fig:quantum-teleportation}
\end{figure}

\autoref{fig:quantum-teleportation} illustrates a circuit realizing quantum teleportation. 
Here, the ancillary qubits \(q_1\) and \(q_2\) are both initialized in basis state \(\ket{0}\) and will be utilized as transportation channel. 
The channel is established by setting \(q_1\) and \(q_2\) into a Bell~state~\cite{bell1964einstein}, i.e.,~entangling both qubits.
After establishing the channel, \(q_1\) and \(q_2\) can be physically separated as indicated by the dashed line in the figure.

With the teleportation channel set up, teleporting the state of \(q_0=\ket{\varphi}\) through the channel is conducted by a Bell measurement. 
The Bell measurement is realized by applying a \(\mathit{CNOT}\) gate with control \(q_0\) and target \(q_1\), followed by a Hadamard gate acting on \(q_0\).
Subsequently, the qubits \(q_0\) and \(q_1\) are measured in the computational basis (i.e.,~\(\ket{0}\) and \(\ket{1}\)).
Since \(q_1\) and \(q_2\) are entangled, the measurement of \(q_1\) also affects the state of \(q_2\).
The qubits \(q_0\) and \(q_1\) together now in one of four the different basis states \(\ket{00}\), \(\ket{01}\), \(\ket{10}\), and \(\ket{11}\)---encoding possible phase- and bit-flip errors in \(q_2\).
The state \(\ket{\varphi}\) can then be established in qubit \(q_2\) by applying---based on the measurement outcome---an \(X\) operation (to correct a bit flip) and/or a \(Z\) operation (to correct a phase flip).
In \autoref{fig:quantum-teleportation}, the classical information (which is classically transferred between the locations of \(q_1\) and \(q_2\)) is denoted by two parallel lines between the measurement and the \(X\) (\(Z\)) gate.

Overall, teleportation requires a constant number of quantum operations and the transfer of only two bits of classical information. 
Since the error rates for \(\mathit{CNOT}\) gates and measurement are similar, the error rates for a single swap (three \(\mathit{CNOT}\) gates) is similar to a quantum teleportation (one \(\mathit{CNOT}\) gate and two measurements).
However, since a teleportation measures qubits \(q_0\) and \(q_1\), they are in a basis state afterwards---the previous transportation channel is destroyed.

Quantum teleportation as introduced above allows to move the state of a qubit over arbitrarily large distances (assuming a transportation channel is established) using rather few quantum operations---making it interesting for the mapping problem described in \autoref{subsec:quantum-circuits}.
In the next section, we show that current mapping approaches can be easily extended to additionally exploit quantum teleportation (which natively fits into NISQ computers)
when mapping quantum circuits to physical devices---allowing for cheaper mapped circuits and a broader search space when aiming for an optimal solution.

\subsection{Quantum Teleportation in Mapping}

This section shows how quantum teleportation natively fits into quantum circuit mapping using swapping. 
As stated above, exploiting quantum teleportation requires

\begin{itemize}
	\item pairs of ancillary qubits that serve as Bell pairs for the quantum transportation channels,
	\item moving the ancillaries away from each other (in order to teleport a state over a \enquote{large} distance),
	\item Bell measurement, and
	\item transferring the classical measurement outcome to correct bit- and phase-flips in the teleported state.	
\end{itemize}

In the following, we show that all these requirements for using quantum teleportation are inherently satisfied when mapping quantum circuits to NISQ devices using swapping. 
Moreover, we show that exploiting teleportations may allow for cheaper mapped circuits. 
Thereby, mapping a quantum circuit to the IBM~Q Tokyo quantum computer (whose coupling map is depicted in \autoref{fig:ibm-tokyo}) serves as example.

Since each quantum computer is composed of a fixed number of $m$ physical qubits (e.g.,~\(m=20\) for IBM Q Tokyo), $m-n$ ancillary qubits are available when mapping an $n$-qubit quantum circuit to the target device. This allows for \(\lfloor\frac{m-n}{2}\rfloor\) ancillary pairs serving as transportation channels. 
Forming pairs of qubits that are \enquote{connected} in the coupling map, the transportation channel is established by using one Hadamard and one \(\mathit{CNOT}\) gate only.

\begin{example}
Consider IBM~Q Tokyo's coupling map depicted in \autoref{fig:ibm-tokyo} and assume that a 18-qubit circuit shall be mapped to this device---allowing for one pair of ancillary qubits \(t_0\) and \(t_1\) to build up a transportation channel. 
Mapping \(t_0\) and \(t_1\) to the adjacent physical qubits \(Q_{17}\) and \(Q_{12}\) allows to establish the transportation channel by applying a Hadamard gate to \(Q_{12}\) and a CNOT gate with control qubit \(Q_{12}\) and target qubit \(Q_{17}\). 
The remaining physical qubits are used to initially map the logical qubits of the quantum circuit.
\end{example}

After establishing the transportation channels, the swapping approach is conducted as usual. 
Since this necessarily moves the logical qubits on the physical qubits of the architecture to satisfy the coupling constraint, the ancillary qubits are moved around as well---potentially increasing the distance between them.

\begin{figure}
	\centering
	\resizebox{\linewidth}{!}{\begin{tikzpicture}[thick, qubit/.style={draw, circle, minimum size=2.75em, fill=white}]
		\matrix (tokyo) [matrix of math nodes, ampersand replacement=\&, every node/.append style={qubit, draw=lightgray, text=lightgray}, row sep=0.5cm, column sep=1cm] {
			Q_0 \& Q_1 \& |[thick, densely dashed, draw=black, text=black]| Q_2 \& |[ultra thick, draw=black, text=black]| Q_3 \& Q_4 \\
			Q_5 \& Q_6 \& Q_7 \& Q_8 \& Q_9 \\
			Q_{10} \& Q_{11} \& Q_{12} \& Q_{13} \& Q_{14} \\
			Q_{15} \& |[ultra thick, draw=black, text=black]| Q_{16} \& |[thick, dashed, draw=black, text=black]| Q_{17} \& Q_{18} \& Q_{19} \\
		};
		\begin{scope}[on background layer, every path/.style={thick, lightgray}]
			\draw (tokyo-1-1) -- (tokyo-1-5);
			\draw (tokyo-2-1) -- (tokyo-2-5);
			\draw (tokyo-3-1) -- (tokyo-3-5);
			\draw (tokyo-4-1) -- (tokyo-4-5);
			
			\draw (tokyo-1-1) -- (tokyo-4-1);
			\draw (tokyo-1-2) -- (tokyo-4-2);
			\draw (tokyo-1-3) -- (tokyo-4-3);
			\draw (tokyo-1-4) -- (tokyo-4-4);
			\draw (tokyo-1-5) -- (tokyo-4-5);
			
			\draw (tokyo-1-2) -- (tokyo-4-5);
			\draw (tokyo-1-3) -- (tokyo-3-1);
			\draw (tokyo-1-4) -- (tokyo-2-5);
			\draw (tokyo-1-5) -- (tokyo-4-2);
			\draw (tokyo-2-1) -- (tokyo-4-3);
			\draw (tokyo-3-5) -- (tokyo-4-4);
		\end{scope}
		\draw[dashed, -Latex, every edge/.append style={black, ultra thick}] (tokyo-4-2) 
			edge node[midway,below,sloped,font=\scriptsize]{1\textsuperscript{st} \(\mathit{SWAP}\)} (tokyo-3-3);
		\draw[dashed, -Latex, every edge/.append style={black, ultra thick}] (tokyo-3-3) 
			edge node[midway,below,sloped,font=\scriptsize]{2\textsuperscript{nd} \(\mathit{SWAP}\)} (tokyo-2-4);
		\draw[ultra thick, dotted,-Latex] (tokyo-4-2) 
			edge[black, bend left=15] node[midway,below,sloped,font=\scriptsize] {teleportation} (tokyo-1-3);
	\end{tikzpicture}}
	\vspace{-2em}
	\caption{Teleportation vs. swapping for \(\mathit{CNOT}(Q_3, Q_{16})\)}
	\label{fig:teleportation-example}
\end{figure}
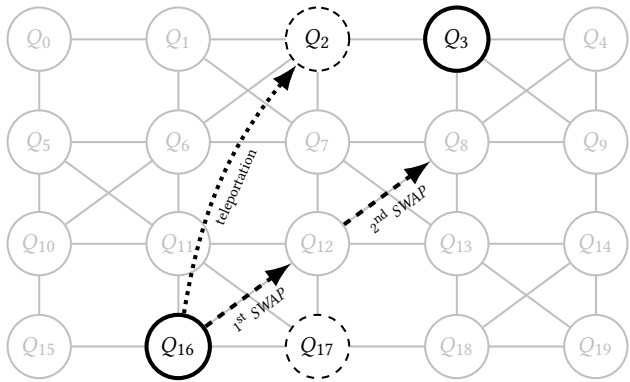

\addtocounter{example}{-1}
\begin{example}[continued]
	Assume that the swapping approach requires to apply swap gates that first swap the state of the physical qubits \(Q_{12}\) and \(Q_7\) and, afterwards, to swap the state of \(Q_7\) and \(Q_2\). 
	This moves the ancillary qubits \(t_0\) and \(t_1\) away from each other and they are now mapped to the physical qubits \(Q_{17}\) and \(Q_2\), respectively, as illustrated by dashed nodes in \autoref{fig:teleportation-example}.
\end{example}

Having a transportation channel where the corresponding logical qubits are mapped to the physical qubits $Q_i$ and $Q_j$ allows to teleport the state of any physical qubit (i.e.,~the logical qubit mapped to this physical qubit) which is adjacent to~\(Q_i\) to~\(Q_j\) (and vice versa) by conducting a Bell measurement. 
Since the required quantum operations (i.e.,~\(\mathit{CNOT}\), \(H\), and measurement in the computational basis) are available on current quantum computers, this step also fits natively into existing mapping algorithms. 
Furthermore, depending on the \emph{binary} results of the measurements, the gates \(X\) and/or \(Z\) are applied to correct for possible bit- and phase-flip errors in the prepared entanglement. 
Notably, the last two gates are single-qubit gates, since the classical computer controlling the quantum computer can unambiguously decide whether to apply the gates or not.
 
Thinking in terms of coupling maps, possible teleportations can be represented by additional \emph{virtual} edges from qubits adjacent of \(Q_i\) (\(Q_j\)) to \(Q_j\) (\(Q_i\)) that are temporarily added to the coupling map. This further increases the search space when mapping quantum circuits to physical devices---allowing for cheaper mapped circuits in the best case. More precisely, the additional virtual edges in the coupling map offer new potential for moving around the logical qubits to satisfy the coupling constraints.

\addtocounter{example}{-1}
\begin{example}[Continued]
In \autoref{fig:teleportation-example}, the ancillary qubits \(t_0\) and \(t_1\) (which establish a transportation channel) are mapped to \(Q_2\) and \(Q_{17}\).
Using this transportation channel, the state of each qubit adjacent to \(Q_2\) can be teleported to the other \(Q_{17}\) (and vice versa).
More precisely, the state of either \(Q_{11}\), \(Q_{12}\), \(Q_{16}\), or \(Q_{18}\) can be teleported to \(Q_2\); or the state of either \(Q_{1}\), \(Q_{3}\), \(Q_{6}\), or \(Q_{7}\) can be teleported to \(Q_{17}\). 
This can be thought of eight additional virtual edges in the coupling map.

Assume that a \(\mathit{CNOT}\) operation shall be applied between logical qubits mapped \(Q_3\) and \(Q_{16}\) (bold nodes in \autoref{fig:teleportation-example}).
The coupling constraints may be satisfied by applying swapping twice as sketched by the dashed lines between \((Q_{16}, Q_{12})\) and \((Q_{12}, Q_8)\).
However, the coupling constraint can also be satisfied by teleporting the state of \(Q_{16}\) to \(Q_2\) (along the virtual edge as sketched by the dotted line in \autoref{fig:quantum-teleportation}).

While satisfying the coupling constraints via swapping requires six \(\mathit{CNOT}\) operations, using teleportation requires only two \(\mathit{CNOT}\)s, two Hadamard gates, two measurements, and up to two further single-qubit gates to correct errors in the teleported state depending on the measurement outcome. 
Depending on the \emph{cost} of the individual operations, teleportation has the potential to be cheaper than exclusively swapping.
\end{example}

The transportation channel vanishes after teleporting the state of a qubit due to the Bell measurement. 
However, since the measured physical qubits are connected in the coupling map, they can be easily entangled again---resulting in a new transportation channel that might be used later in the mapping procedure.
Before entangling the qubits (which are in a basis state, see \autoref{fig:quantum-teleportation}), they have to be reset to basis state zero (by applying an \(X\) gate in case that measurement outcome was basis state \(\ket{1}\)).\footnote{Gates establishing transportation channels that are never used in the mapping procedure can be easily removed from the circuit by according post-mapping optimizations.}

\addtocounter{example}{-1}
\begin{example}[continued]
	After the teleportation, \(Q_{16}\) and \(Q_{17}\) are in a basis state, i.e.,~either~\ket{0} or~\ket{1}. In the latter case, applying an \(X\)~operation ensures the basis state~\ket{0}.
	Since the measured qubits are adjacent and in state~\ket{0}, they can be easily re-entangled as shown in \autoref{fig:quantum-teleportation}.
	As the mapping process continues, \(\mathit{SWAP}\) operations move the state of these newly entangled qubits around, likely increasing their distance and, hence, the probability that they might be in a suitable position later for the next teleportation.
\end{example}

The proposed approach leads to a blueprint for exploiting quantum teleportation during quantum circuit mapping. 
As described above, this fits into swapping strategies and can be applied when mapping circuits to current NISQ devices. 
In general, additionally using teleportation cannot worsen mapping results since, in the worst case, no teleportations are conducted at all. 
However, finding a mapping approach that utilizes teleportation to reduce the cost of the mapped circuit is nontrivial and requires to quantify the cost of a teleportation compared to the cost of a \(\mathit{SWAP}\) operation.
A first case study showcasing the potential is presented in the following section.

\section{Case Study}
\label{sec:case-study}

This section provides a case study to demonstrate the benefit of additionally exploiting quantum teleportation in mapping approaches based on swapping. 
To this end, the IBM Q Tokyo device (see \autoref{fig:ibm-tokyo}) serves as target architecture and the benchmarks consist of functions from RevLib~\cite{WGT+:2008} as well as quantum functionality.

For the case study, we incorporated the idea of quantum teleportation into the A$^*$-based mapping algorithm described in~\cite{DBLP:journals/tcad/ZulehnerPW19} (which has been downloaded from \url{https://iic.jku.at/eda/research/ibm_qx_mapping/}).
The general idea of~\cite{DBLP:journals/tcad/ZulehnerPW19} is to first partition the circuit to be mapped into layers of quantum gates acting on disjoint qubits. Starting from an initial mapping of the logical qubits to the physical ones (which allows applying all gates from the first layer), the cheapest permutation of the mapping (and the corresponding sequence of \(\mathit{SWAP}\) operations) is determined before each layer such that all gates of the layer can be applied.\footnote{Look-ahead strategies are utilized to aim for a global minimum rather than local ones. Details are provided in~\cite{DBLP:journals/tcad/ZulehnerPW19}.}
Based on that, teleportation effectively adds virtual edges, which are considered in the search as well as the predefined edges of the coupling map.
Hence, the virtual edges further increase the search space traversed by the A$^*$-algorithm, potentially allowing to determine a more cost efficient mapping.

In the following, we denote the original A$^*$-based mapping algorithm (solely utilizing swapping) as $M_{\textit{SWAP}}$, while we denote the proposed algorithm that also utilizes teleportations as $M_{\textit{SWAP+TEL}}$.
Conducting an evaluation of these two algorithms is nontrivial since calling the A$^*$-algorithm before each layer results in many local choices that may heavily influence the cost of the resulting mapped circuit.
To overcome this issue, we randomly chose 50 initial mappings (satisfying the coupling constraint of the gates in the first layer) for each benchmark, and determined the obtained minimum overhead.
We utilized quantum circuits and RevLib circuits considered for evaluating previous mapping algorithms as benchmarks~\cite{WGT+:2008}.

\autoref{tab:results} summarizes the obtained results.
For each benchmark, we list its name, number of qubits, as well as the number of gates.
The fourth column lists the number of \(\mathit{SWAP}\) gates added by $M_{\textit{SWAP}}$ during the mapping. Note that we list only the best result  (out of 50 runs) for each benchmark. 
The next two columns list the number of \(\mathit{SWAP}\) operations and the number of teleportations added by $M_{\textit{SWAP+TEL}}$. 
The runtimes have been omitted since each mapping has been determined in less than 10 seconds, often even in less than a second.
The mapping was performed on a system running GNU/Linux with an Intel i7-7700K CPU (\SI{4.2}{\giga\hertz}) and \SI{32}{\gibi\byte} main memory.

In order to obtain the respectively required overhead, we applied a cost function that weighs \(\mathit{SWAP}\) operations compared to teleportations. 
We considered the following\footnote{Both cost functions resulted in the same best solution for each benchmark.}:
\begin{itemize}
	\item A function assigning equal costs to \(\mathit{SWAP}\) and teleportation.
	\item A function based on gate error-rates used by IBM to evaluate mapping algorithms in the IBM Qiskit Developer Challenge 2018, where a cost of 10 is assigned to each \(\mathit{CNOT}\) and a cost of 1 is assigned to (almost) each \mbox{single-qubit} gate. 
	We extended IBM's cost function by assuming also costs of 10 for a measurement since their error rate is similar to \(\mathit{CNOT}\)~gates.
\end{itemize}
The last two columns of \autoref{tab:results} list the relative cost difference of $M_\textit{SWAP+TEL}$ compared to $M_\textit{SWAP}$ according to these cost functions.

\begin{table}[!tpb]
	\caption{Case Study on Swapping with Teleportation}
	\label{tab:results}
	\vspace{-1em}
	\centering
	\footnotesize
	\setlength{\tabcolsep}{0.05cm}
	\resizebox{\linewidth}{!}{\begin{tabular}{lr@{\hskip 8pt}r@{\hskip 16pt}r@{\hskip 16pt}rr@{\hskip 16pt}rr}
		\toprule
		\multicolumn{3}{c}{benchmark\hspace*{8pt}} & $M_{\textit{SWAP}}$ & \multicolumn{2}{c}{$M_{\textit{SWAP+TEL}}$\hspace*{14pt}} & \multicolumn{2}{c}{rel. cost}\\ 
		\cmidrule(r{14pt}){1-3}\cmidrule(r{14pt}){4-4}\cmidrule(r{14pt}){5-6}\cmidrule{7-8}
		name & qubits & gates & \#SWAP & \#SWAP & \#TEL. & equal & IBM \\ \midrule
				\begin{filecontents*}{results.csv}
name,qubits,gates,depth,SWAPS1,SWAPS2,TELEPORTATIONS,,
ex3\_229,6,403,226,8,6,1,0.88,0.93
mini\_alu\_305,10,173,69,17,11,1,0.71,0.73
rd73\_140,10,230,92,23,21,1,0.96,0.98
sym9\_148,10,21\thinspace504,12\thinspace087,1\thinspace657,1\thinspace369,19,0.84,0.84
life\_238,11,22\thinspace445,12\thinspace511,2\thinspace638,2\thinspace573,33,0.99,0.99
z4\_268,11,3\thinspace073,1\thinspace644,329,311,7,0.97,0.98
cm152a\_212,12,1\thinspace221,684,116,86,7,0.80,0.83
sym9\_146,12,328,127,40,34,3,0.93,0.96
dist\_223,13,38\thinspace046,19\thinspace694,4\thinspace684,4\thinspace496,57,0.97,0.98
radd\_250,13,3\thinspace213,1\thinspace781,344,303,7,0.90,0.91
rd53\_311,13,275,124,36,34,1,0.97,0.98
root\_255,13,17\thinspace159,8\thinspace835,2\thinspace032,1\thinspace992,11,0.99,0.99
squar5\_261,13,1\thinspace993,1\thinspace049,187,178,4,0.97,0.98
sym10\_262,13,64\thinspace283,35\thinspace572,8\thinspace316,7\thinspace745,191,0.95,0.96
0410184\_169,14,211,104,30,24,4,0.93,0.99
cm85a\_209,14,11\thinspace414,6\thinspace374,1\thinspace180,1\thinspace157,11,0.99,0.99
pm1\_249,14,1\thinspace776,940,148,111,3,0.77,0.78
sym6\_316,14,270,135,34,29,1,0.88,0.90
co14\_215,15,17\thinspace936,8\thinspace570,2\thinspace529,2\thinspace401,32,0.96,0.97
dc2\_222,15,9\thinspace462,5\thinspace242,1\thinspace119,1\thinspace078,9,0.97,0.97
ham15\_107,15,8\thinspace763,4\thinspace819,965,948,9,0.99,1.00
misex1\_241,15,4\thinspace813,2\thinspace676,433,372,8,0.88,0.89
rd84\_142,15,343,110,38,32,1,0.87,0.88
square\_root\_7,15,7\thinspace630,3\thinspace847,904,844,10,0.94,0.95
cnt3-5\_179,16,175,61,30,25,1,0.87,0.88
ising\_model\_16,16,786,71,8,3,2,0.63,0.73
qft\_16,16,512,105,88,81,5,0.98,1.00
\end{filecontents*}
\csvreader[
		late after line=\\,
		late after last line=\\,
		]{results.csv}
		{1=\Name,2=\Qubits, 3=\Gates, 4=\Depth, 5=\Swaps, 6=\SwapsT, 7=\Teleportations, 8=\CostA, 9=\CostB}
		{\Name & \Qubits & \Gates & \Swaps & \SwapsT & \Teleportations & \CostA & \CostB}
		\bottomrule
	\end{tabular}}
	
	\vspace{4pt}
	{\raggedright
	equal: Assuming equal costs for \(\mathit{SWAP}\) and teleportation \\
	IBM: Using the cost function motivated by \emph{IBM Q Developer Challenge 2018}\\
	Each run of the benchmark was completed in less than \SI{10}{\second}.

	}
\end{table}

The results show that, even if the considered architecture is rather small, utilizing teleportation during mapping quantum circuits indeed significantly reduces the overhead in many cases---independently of what cost function (equal or IBM) is applied. 
In the best cases, i.e.,~for benchmarks mini\_alu\_305, cm152a\_212, or pm1\_249, the overhead can be reduced by \SI{20}{\percent} and up to almost \SI{30}{\percent}---an impressive improvement considering that the distances in IBM Q Tokyo architecture are still rather small. Motivated by these numbers, we are certain that, with the emergence of larger quantum architectures and, hence, larger distances to cover, teleportation is going to become more effective with growing architectures. 

\section{Conclusions}
\label{sec:conclusions}

The state of the art for mapping quantum circuits to actual architectures is utilizing classical thinking by only swapping moving states along the static coupling map.
In this work, we showed that additionally exploiting quantum teleportation as possibility provided by quantum mechanics, enables further options for moving qubit states in order to satisfy the coupling constraint.
In fact, the proposed scheme allows to connect qubits over arbitrary long distances with a constant overhead.
The constant costs will will lead to a greater impact when larger quantum architectures become accessible (such as the ones announced by IBM~\cite{ibm2020roadmap} and Google~\cite{arute2019quantum}).
Together with further reduced error rates for gates and measurements as the hardware matures, the full potential of quantum teleportation in quantum circuit mapping will be available.
But even for architectures such as available today, such as the 20 qubit IBM~Q~Tokyo architecture, significant improvements can be achieved now in many cases. 
Because of that, we are certain that utilizing quantum teleportation is going to become more effective with growing architectures.

\section*{Acknowledgments}
This work has partially been supported by the LIT Secure and Correct Systems Lab funded by the State of Upper Austria as well as by BMK, BMDW, and the State of Upper Austria in the frame of the COMET Programme managed by FFG.

\let\oldbibliography\thebibliography
\renewcommand{\thebibliography}[1]{%
  \oldbibliography{#1}%
  \setlength{\itemsep}{1pt plus 1pt minus 1pt}%
}
\bibliographystyle{ieeetr}
\bibliography{lit_header,lit_references}
\end{document}